\documentclass[conference]{IEEEtran}
\IEEEoverridecommandlockouts

\usepackage{cite}
\usepackage{amsmath,amssymb,amsfonts,fixmath}
\usepackage{algorithm}
\usepackage{subcaption}
\usepackage{graphicx}
\usepackage{textcomp}
\usepackage{xcolor}
\def\BibTeX{{\rm B\kern-.05em{\sc i\kern-.025em b}\kern-.08em
    T\kern-.1667em\lower.7ex\hbox{E}\kern-.125emX}}

\usepackage{pgfplots}
\pgfplotsset{compat=newest}
\pgfplotsset{plot coordinates/math parser=false}
\newlength\figureheight
\newlength\figurewidth
\usepgfplotslibrary{patchplots,polar,fillbetween}

\usetikzlibrary{spy,shapes,arrows,positioning,plotmarks,shadows,calc,matrix,fit,patterns}
\usepackage{grffile}
\usepackage{tikzscale}
\usepackage{siunitx}
\DeclareSIUnit{\belmilliwatt}{Bm}
\DeclareSIUnit{\dBm}{\deci\belmilliwatt}
\usepackage{algpseudocode}
\algrenewcommand\textproc{}
\newcommand{\var}{\texttt}

\newcommand\copyrightnotice{%
	\begin{tikzpicture}[remember picture,overlay]
		\node[anchor=north,yshift=-10pt] at (current page.north) {\fbox{\parbox{\dimexpr\textwidth-\fboxsep-\fboxrule\relax}{\footnotesize \textcopyright 2022 IEEE.  Personal use of this material is permitted. Permission from IEEE must be obtained for all other uses, in any current or future media, including reprinting/republishing this material for advertising or promotional purposes, creating new collective works, for resale or redistribution to servers or lists, or reuse of any copyrighted	component of this work in other works.}}};
	\end{tikzpicture}
}

\begin{document}

\title{Full-Duplex meets Reconfigurable Surfaces: RIS-assisted SIC for Full-Duplex Radios\\
\thanks{This work was funded by the Federal Ministry of Education and Research (BMBF) of the Federal Republic of Germany (F\"orderkennzeichen 16KIS1152, mINDFUL and 16KIS1235, MetaSEC).}
}

\author{Simon Tewes$^{\ast }$, Markus Heinrichs$^{\dagger }$, Paul Staat$^{\ddagger }$, Rainer Kronberger$^{\dagger}$, Aydin Sezgin$^{\ast }$\\
		$^{\ast }$Chair of Digital Communication Systems, Ruhr-University Bochum, Bochum, Germany,\\
		$^{\dagger}$High Frequency Laboratory, TH Cologne - University of Applied Sciences, Cologne, Germany\\
		$^{\ddagger }$Max Planck Institute for Security and Privacy, Bochum, Germany,\\
    	 \{simon.tewes, aydin.sezgin\}@rub.de,\{markus.heinrichs, rainer.kronberger\}@th-koeln.de, paul.staat@mpi-sp.org}
		
\maketitle
\copyrightnotice
\begin{abstract}
Reconfigurable intelligent surfaces~(RIS) are a key enabler of various new applications in 6G smart radio environments. By utilizing an RIS prototype system, this paper aims to enhance self-interference~(SI) cancellation for in-band full-duplex~(FD) communication systems. SI suppression is a crucial requirement for FD communication as the SI severely limits the performance of a node by shadowing the received signal from a distant node with its own transmit signal. To this end, we propose to assist SI cancellation by exploiting an RIS to form a suitable cancellation signal in the analog domain.

Building upon a 256-element RIS prototype, we present results of RIS-assisted SI cancellation from a practical testbed. Given an initial analog isolation of \SI{44}{\decibel} provided by the antenna design, we are able to cancel the leaked signal by an additional \SI{59}{\decibel} in the narrowband case, resulting in an overall SI suppression of \SI{103}{\decibel} without additional digital cancellation. The presented case study shows promising performance to build an FD communication system on this foundation.

\end{abstract}


\section{Introduction}
The ever-growing communication demand is expected to reach 5.3 billion total internet users in 2023 according to the Cisco annual internet report (2018-2023) \cite{cisco}. With this evolution of wireless data transfer and rapidly increasing numbers of devices, e.g., through the internet of things, new communication standards are evolving. With the ongoing standardization of 5G, a new generation of communication systems is currently rolled out. The research attention, therefore, is already drawn to the next generation wireless communications, commonly referred to as 6G. Among many other promising applications and new techniques, research on 6G investigates the use of reconfigurable intelligent surfaces~(RIS). The RIS is a synthetic surface able to passively manipulate the reflected signal depending on an electrical reconfiguration of the elements. For the first time, this promising concept allows to deviate from the paradigm of a determined, passive communication channel and rather allows to optimize the channel for a specific application or user. These innovative features qualify the RIS to be a part of an upcoming communication standard \cite{zhao2019survey}.

In this paper, we focus on applying the RIS in an in-band full-duplex~(FD) communication system to enhance self-interference~(SI) cancellation. FD describes the simultaneous transmission and reception on the same carrier frequency. FD is a key enabler for secure key generation \cite{PStaat} and exchange over the air \cite{SecureMIMO}, jamming resistant communications \cite{AntiJamming} and data throughput maximization~\cite{zhao2019survey}. However, current state-of-the-art FD implementations suffer from high implementation cost and high integration effort of SI cancellation techniques in the analog domain.

Addressing this issue, we propose a novel concept which utilizes an RIS to shape a phase-inverted cancellation signal directed to the receive antenna of an FD-node, adaptively reducing the SI by cancellation. This enables a relaxed transceiver design, which reduces implementation costs and takes advantage of the RIS infrastructure that likely will be deployed in 6G networks. Our approach provides all the advantages of  FD connectivity while reducing UE complexity, potentially even allowing conventional transceiver architectures to be used for FD communication.

\subsection{Prior work}
In the past several concepts for SI cancellation in FD communication systems have been shown \cite{Katti1, Katti2, Duarte1, Duarte2}. In this section, we briefly introduce the core concepts and current state-of-the-art.

In \cite{FDradios}, Bharadia et al. have shown a customized SI cancellation printed circuit board (PCB) containing a number of fixed delays and variable attenuators. These delayed and attenuated signals bypass an external RF circulator that combines the transmitter and receiver to a single antenna. A control algorithm is used to parameterize the attenuators in order to generate a cancellation signal combating the leakage in the setup. Additionally, a digital cancellation step is employed to eliminate residual leakage. Even though, this setup shows good performance in the presented scenarios, it involves a high number of analog RF components, that, on the one hand, are pricey and, on the other hand, may not be easily integrated into SoCs. Our approach does not rely on external RF components, but rather uses the RIS infrastructure of future radio network environments reducing user equipment~(UE) costs significantly.

In \cite{Vogt}, Vogt et al. have shown a scheme, where an external circulator was used to combine the transmitter and receiver on a single antenna. Additionally, an auxiliary transmitter was used to eliminate the residual SI in the analog domain by adding a cancellation signal with an RF combiner to it. Furthermore, a digital cancellation step is employed. This setup, however, requires an external circulator, combiner, and an auxiliary transmitter adding significant cost to the design which we spare in our proposed approach.

In SoftNull \cite{Everett}, the authors present a MIMO approach to FD SI cancellation. They have shown an antenna array of 72 antennas, where some of the antennas are used for transmitting and receiving, while others are used for cancellation. In their paper, the authors exploit knowledge of the channel between the antennas $H_{\text{Self}}$ to actively use a subset of antennas to reduce SI on selected antennas of the array. This precoding procedure enables FD on some antennas, however sacrificing some transceiver chains and antennas in the array for cancellation. Furthermore, the array is physically large, limiting the use on mobile UEs. In our approach, we only need a single transmit and a single receive antenna and no additional transceivers.

All of these previously outlined concepts use specialized RF hardware \cite{FDradios}, \cite{Vogt}, or large antenna arrays \cite{Everett}. This is why we propose to utilize existing RISs in future 6G networks, relaxing the requirements to specialized hardware on the UE side. This can significantly reduce costs and save complexity at the UE. We propose to turn a regular half-duplex transceiver with two antennas to a full-duplex transceiver temporarily, when needed, e.g., for secret key-exchange. By using channel estimation, the RIS is tuned so that the reflected signal cancels the cancels the SI at the UE and thus enables FD communication. 

The remainder of the paper is structured as follows. First, in Section \ref{sec:RISforSI} we introduce our approach to RIS-assisted SI cancellation. Then, we detail the setup of the conducted experimental study and the used hardware. Next, we propose a greedy algorithm for generating a suitable RIS configuration. Finally, in Section \ref{sec:ExpResults} we provide the results of the experiments and present the insights drawn from them.

\section{RIS for analog self-interference cancellation}\label{sec:RISforSI}
The main challenge in in-band FD reception is to cancel the simultaneously transmitted signal of the FD node. If the leakage of the transmitted signal into the receive-chain of the transceiver is high, the utilizable dynamic range of the analog to digital converter in the receiver is reduced drastically. This leakage either  overwhelms the received signal of the distant node or the degraded SNR reduces data throughput significantly.

Here, our approach is to reduce the SI by utilizing an RIS to form a suitable cancellation signal combating the leakage. An FD node with two antennas is considered, one antenna to transmit and one to receive, respectively. Utilizing two spatially separated antennas on an FD node yields an amount of initial analog isolation between transmitter and receiver chain due to antenna design.
This setup results in a leaked signal to the receiver, with a signal power $P_{leaked}$ given by
\begin{align}
	P_{\text{leaked}}=P_{\text{tx}}-\alpha_{\text{iso}},
\end{align}
where $\alpha_{\text{iso}}$ (in dB) is the initial analog isolation representing the total leakage between the transmit and receive path, including path loss, antenna gains, cable losses, and internal coupling. $P_{\text{tx}}$ (in dBm) is the transmit power of the FD node. The received signal power stemming from the distant FD node at the receive antenna is attenuated by approximately the free space path loss between both FD nodes.

\begin{figure}
	\centering
	\includegraphics[width=\linewidth]{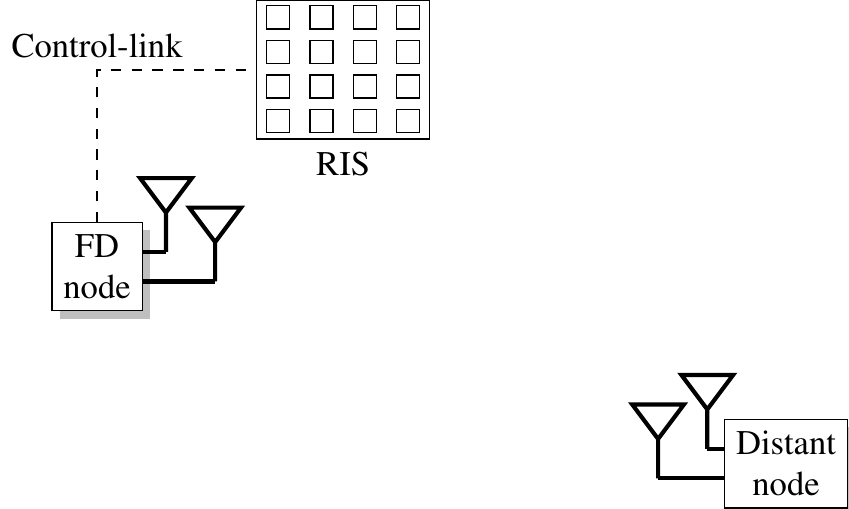}
	\caption{System Overview}
	\label{fig:SystemModel}
\end{figure}
The received signal power (also in dBm) is given by
\begin{align}
	\begin{split}	
		P_{\text{rx}}=P_{\text{tx}}-20 	\log_{10}(d)-20 \log_{10}(f)\\- 20 \log_{10}\left(\frac{4\pi}{c_0}\right)		+G_t+G_r,
	\end{split}
\end{align}
where $d$ is the distance between both FD nodes, $f$ is the carrier frequency and $G_t$ and $G_r$ (in dBi) are the antenna gains, respectively.
The received power $P_{\text{rx}}$ is usually orders of magnitude lower than the received power due to SI,~$P_{\text{leaked}}$.
That is why we propose to utilize the RIS to form a phase-shifted cancellation signal to further reduce $P_{\text{leaked}}$.
The reflected signal of the RIS $r$ can be modeled as \cite{WCTRIS}
\begin{align}
r=\Biggl[\sum_{i=1}^{N}h_ie^{j\phi_i}g_i\Biggr]x,
\end{align}
where $\phi_i$ is the phase shift introduced by the $i$th element of the RIS, $h_i$ and $g_i$ are the channels from the transmitter to the $i$-th element of the RIS and from the $i$-th element to the receiver, respectively. $N$ is the number of total elements of the RIS and $x$ is the transmit signal.
From a link budget perspective, given a perfect RIS cancellation signal and neglecting non-linearities, this leads to 
\begin{align}\label{equ:Psi}
	P_{\text{si}}=P_{\text{tx}}-\alpha_{\text{iso}}-P_{\text{RIS}},
\end{align}
where $P_{\text{si}}$ is the residual self-interference power and $P_{\text{RIS}}$ is the power reflected from the RIS to the receive antenna (both in dBm).

The goal of the case study presented in this paper is to determine an RIS parameter setting, at which $P_{\text{leaked}}$ and $P_{\text{RIS}}$ cancel each other out as much as possible. To achieve maximum cancellation, the reflected signal must not only be phase inverted with respect to the leaked signal, but also have the same amplitude. If the power of the reflected signal is higher than the power of the leaked signal, then the reflected signal overshoots and the minimum cannot be reached. These combined requirements make finding a good RIS setting very challenging. In the remainder of the paper, we present results gathered from a prototypical lab setup.
\section{Experimental study setup}
In this section, we describe the RIS, the transceiver system, the methodology of RIS configuration, and the experimental setup. 

\subsection{The RIS prototype}
The RIS used is a $16 \times 16$ array of unit cells with identical structure embedded on a PCB. The resonance frequency of each individual unit cell is binary-switchable, resulting in different reflection coefficients with different phase for the two switching states. The unit cell consists of a rectangular patch reflector and a contiguous ground plane. The length of the patch is in the order of half a wavelength and thus defines the resonance frequency of the unloaded patch reflector. A via connects the patch on the top side to the anode of a PIN diode on the bottom side of the RIS. The cathode of the PIN diode is connected to the ground plane. The via is not positioned in the center of the patch, leading to an altered electrical length of the patch reflector when the PIN diode is biased and shorts the via to ground. This results in a binary-switchable resonance frequency of the unit cell.

\begin{figure}
	\centering
	\includegraphics[width=.9\linewidth]{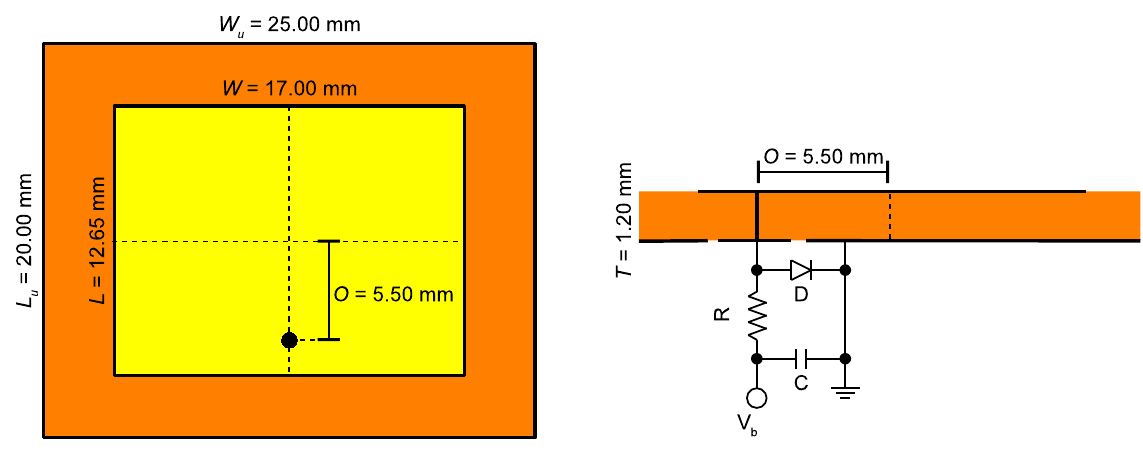}
	\caption{Physical dimensions, and driver circuit of an RIS unit cell.}
	\label{fig:UnitCell}
\end{figure}

The physical dimensions as well as the driver circuitry of the unit cell are shown in Fig.~\ref{fig:UnitCell}. For the substrate, standard low-cost FR4 with a dielectric constant of $\epsilon_r \approx 4.5$ and a dissipation factor of $\tan \delta \approx 0.02$ is used. The PIN diode is of type SMP1320-079LF from manufacturer Skyworks. The biasing resistor has a value of $R = \SI{4.7}{\kilo\ohm}$, resulting in a forward current through the PIN diode of $I_f = \SI{1}{\milli\ampere}$. The decoupling capacitor has a value of $C = \SI{100}{\pico\farad}$. A biasing voltage of $V_b = \SI{5}{\volt}$ or $V_b = \SI{0}{\volt}$ is applied in the ON state or the OFF state, respectively. This unit cell is an evolution of our prior work from \cite{Heinrichs}.

\begin{figure}
	\centering
	\includegraphics[width=.9\linewidth]{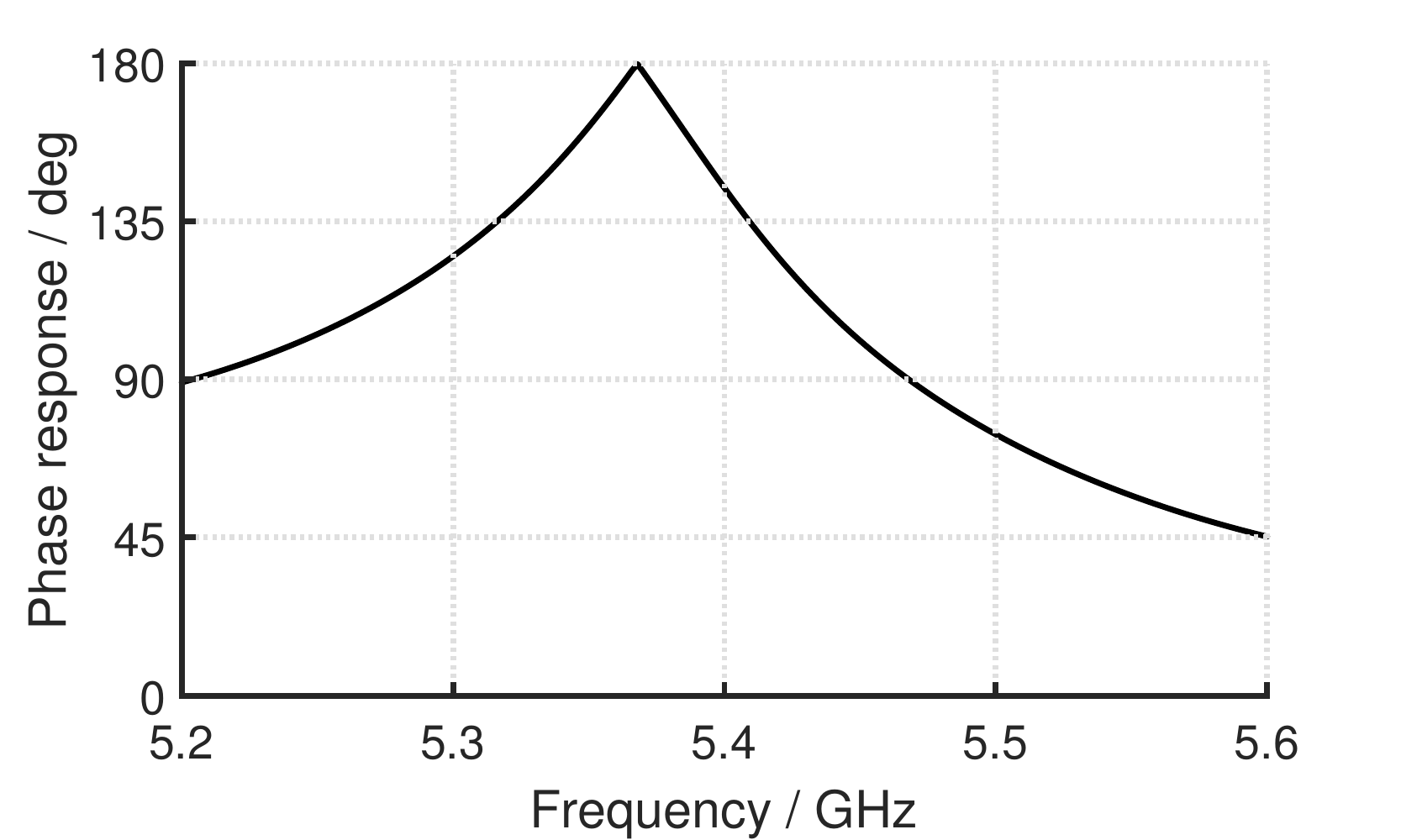}
	\caption{Measured phase response of the RIS.}
	\label{fig:RISPhaseResponse}
\end{figure}

The measured phase response of the RIS is shown in Fig.~\ref{fig:RISPhaseResponse}. This is the phase difference of the surface reflection coefficient between the RIS configured as all-ON and all-OFF states. The measurement was taken under the condition that the incident wave vector and the reflected wave vector are perpendicular to the surface.

\subsection{Transmitter and Receiver}
To verify the effect of the RIS on the SI we use a vector network analyzer (VNA) of type E5071C ENA from Manufacturer Agilent Technologies \cite{agilent} first and later transfer our ground truth measurements from the VNA to software-defined radios (SDRs). The SDRs are Ettus Research X310 \cite{X310} with CBX-120 daughterboards \cite{CBX} suitable for the \SI{5}{\giga \hertz} passband frequency of the RIS. The self-interference due to leakage in the antenna and transceiver assembly is investigated by analysis of the transmission coefficient $s_{21}$ between transmit and receive antenna.

\subsection{Methodology of RIS switching}\label{sec:Methodology}
The $16 \times 16$ RIS used during our experiments offers discrete switching states for all 256~elements. This results in $2^{256}$~possible configuration states. Due to this enormous set of configurations, an exhaustive search of all possible combinations is infeasible. We therefore propose a greedy heuristic to converge to a local optimal RIS setting. The pseudocode of the proposed algorithm is given in Algorithm 1.

\begin{algorithm}
	\caption{Greedy algorithm for RIS setting optimization}
	\begin{algorithmic}
		\renewcommand{\algorithmicrequire}{\textbf{Input:}}
		\renewcommand{\algorithmicensure}{\textbf{Output:}}
		\Require Buffer size $\mathit{B}$, RIS dimension $N$, $f_{\text{tone}}$, termination threshold $t_e$
		\Ensure Best RIS setting and corresponding self-interfence 
		\State 
		
		\\ \textit{Initialisation} :
		\State Termination counter $t_c=0$
		\For {$q = 0$ to $\mathit{B}-1$}
			\State $\mathbold{S}$[q,:,:]=randomBoolean(0.5$\cdot$ones($N_x$,$N_y$))
			\State $\mathbold{i}$[q]=evaluateRISsetting( $\mathbold{S}$[q,:,:])
		\EndFor
		
		\State [$\mathbold{i}$,idx]=sort($\mathbold{i}$) \Comment{Sort according to SI magnitude}
		\State $\mathbold{S}$=$\mathbold{S}$(idx)
		
		\\ \textit{Main Loop} :
		\While{$t_c < t_e$}
		\State $\mathbold{S}_{w}=\mathbold{S}\cdot (1:100)$ \Comment{Linear weighting}
		\State $P=sum(\mathbold{S_w},1)$  \Comment{Sum along first dimension}
		\State $P_\text{norm}=P./\frac{\mathit{B}^2+\mathit{B}}{2}$
		\State $\mathbold{S}_\text{c}$=randomBoolean($P_\text{norm}$)
		\State $I_\text{c}$=evaluateRISsetting($\mathbold{S}_\text{c}$)
		\If {$I(0)<I_\text{c}$} \Comment{New best SI found?}
		\State $t_c=t_c+1$
		\Else
		\State $t_c=0$ 
		\EndIf
		\If {$\mathbold{i}(\mathit{B}-1)>I_\text{c}$} \Comment{Add to buffer}
		\State $\mathbold{i}(\mathit{B}-1)=I_\text{c}$
		\State $S(\mathit{B}-1,:,:)=S_\text{c}$
		\State [$\mathbold{i}$,idx]=sort($\mathbold{i}$)
		\State $\mathbold{S}=\mathbold{S}$(idx)
		\EndIf
		\EndWhile		
		\\ \Return [$\mathbold{S}[0,:,:];\mathbold{i}(0)$] \Comment Return best Setting and SI
		\\ \Function{evaluateRISsetting}{\var{RISsetting}}
		\State Set RIS hardware state to RISsetting
		\State Evaluate SI-Amplitude with VNA or SDR
		\State\Return max(SI-Amplitude)	
		\EndFunction
		
	\end{algorithmic}
\end{algorithm}

We consider two variants of the algorithm. First, the 'Narrowband'-variant, where the algorithm minimizes the SI at a single frequency point, which is \SI{5.385}{\giga\hertz} throughout this paper. Second a variant, that evaluates the SI at multiple equidistand frequency points. The boolean variable 'Narrowband' in algorithm 1 distinguishes both variants. In the remainder, we label the corresponding results 'Narrowband* or with the configured bandwidth '5~MHz' or '10~MHz'
The greedy algorithm utilizes two buffers of size $\mathit{B}$. The buffer $\mathbold{i}_{B\times 1}$ holds the $\mathit{B}$ best SI magnitude readings found so far, the buffer $\mathbold{S}_{B\times N_x\times N_y}$ contains the corresponding RIS configuration.
The greedy search starts with an initialization phase, where $\mathit{B}$ random RIS configurations are iteratively generated and the corresponding SI magnitude for every setting is measured with the VNA. The configurations and SI magnitudes are added to the buffers $\mathbold{i}$ and $\mathbold{S}$.
After acquiring $\mathit{B}$ initial measurements, the buffered data is used to calculate a ratio P for every element, expressing how many times the surface is active.

Therefore, in a loop, first the buffers $\mathbold{i}$ and $\mathbold{S}$ are sorted by means of the ascending SI magnitude. The sorted settings buffer $\mathbold{S}$ is now linearly weighted, where the setting with the lowest SI magnitude in the buffer gets the highest weight $\mathit{B}$ and the setting with the highest SI magnitude in the buffer is weighted with $1$.
The weighted settings buffer $\mathbold{S}_{w}$ is now summed up along the first dimension, and normalized in the range of 0 to 1 by dividing by $\frac{B^2+B}{2}$. The resulting $N_x\times N_y$ dimensional array $P_{norm}$ contains a ratio, how often each element was active with higher weighting on the settings with lower SI magnitude.
The ratios $P_{norm}$ are now used to generate a new configuration for the RIS. With the new setting applied to the surface, the SI magnitude is measured. If the new reading is lower than the highest SI magnitude in the buffer the new reading replaces this reading in both buffers $\mathbold{S}$ and $\mathbold{i}$. The loop repeats until a predefined number of iterations $t_e$, without a new best setting found have passed.

For a narrowband system a single test tone / frequency point is used to evaluate the SI magnitude. To broaden the bandwidth for OFDM systems multiple equidistant frequency points across the desired bandwidth are considered, of which the highest SI magnitude in each loop is used. By always using the maximum over all frequency points we can ensure the SI magnitude to be at least as low or lower across the specified band, as the value our algorithm converges to.

\subsection{Experimental setup:}\label{sec:Setup}

\begin{figure}
	\centering
	\includegraphics[width=.9\linewidth]{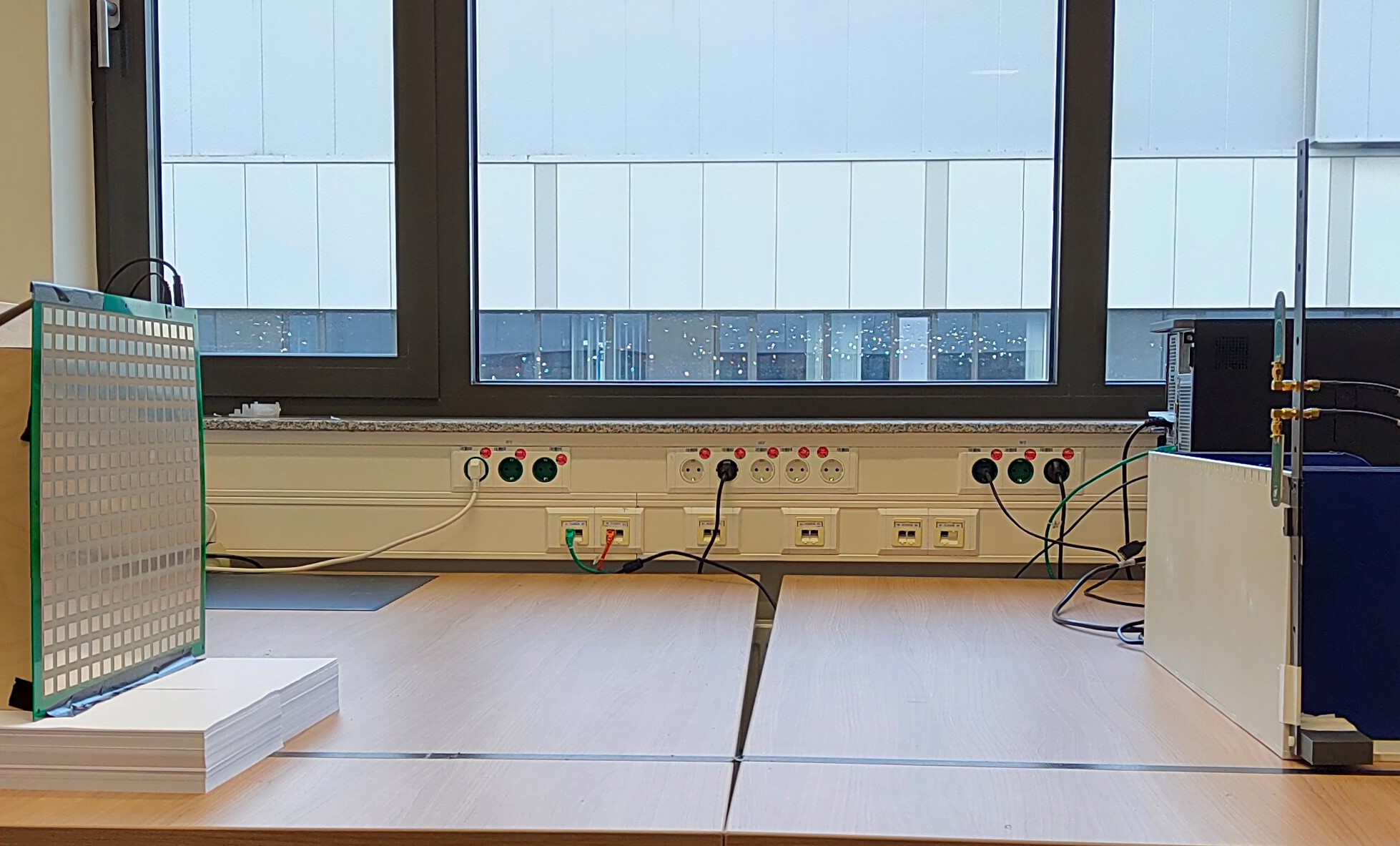}
	\caption{Photography of the RIS setup in a typical office environment. The distance between antennas on the right side and the RIS on the left side is \SI{1}{\meter}.}
	\label{fig:WholeSetup}
\end{figure}

\begin{figure}[h]
	\begin{subfigure}{0.45\linewidth}
		\includegraphics[width=.95\linewidth]{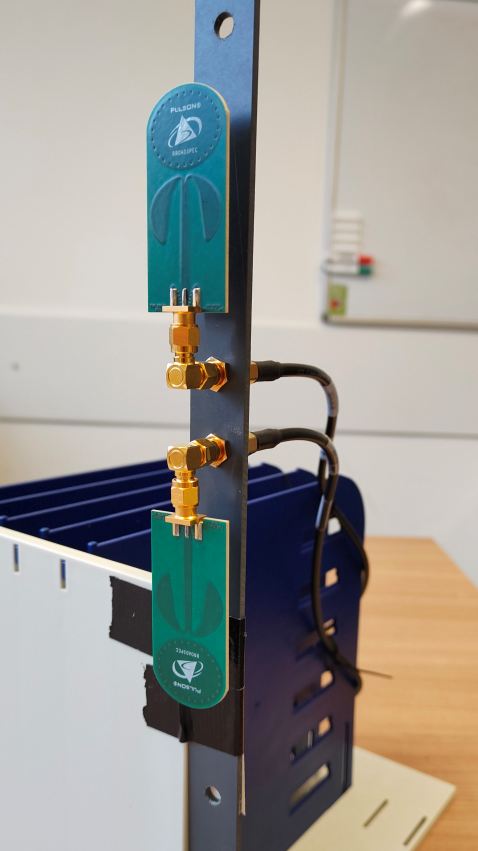}
		\caption{Antenna assembly}
		\label{fig:SplitViewA}
	\end{subfigure} 
	\begin{subfigure}{0.55\linewidth}
		\centering
		\includegraphics[width=.95\linewidth]{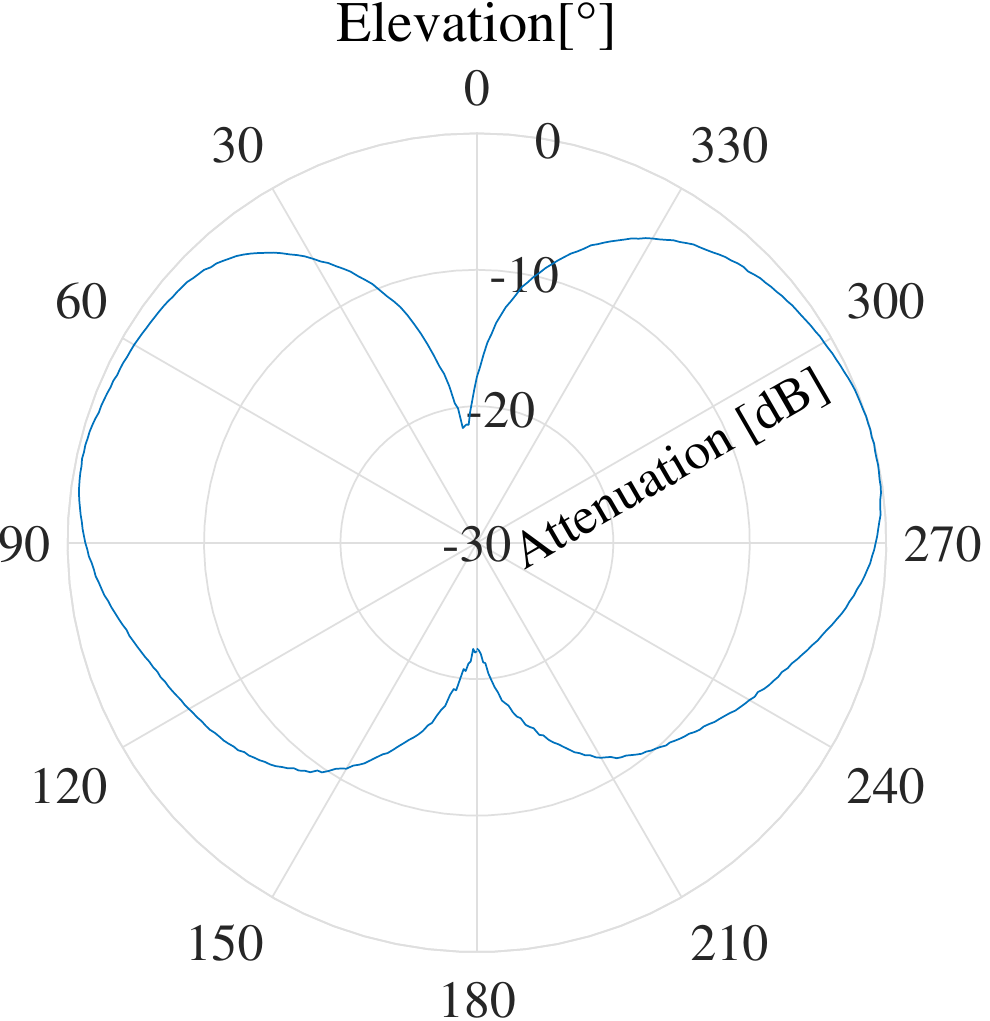}
		\caption{Normalized antenna characteristic at \SI{5.385}{\giga\hertz}}
		\label{fig:SplitViewB}
	\end{subfigure}
	\label{fig:SplitView}
	\caption{On the left, a detailed photography of the antenna arrangement is shown. The distance between the SMA connectors is \SI{25}{\milli\meter}, approx.~$\lambda/2$. On the right, the measured antenna diagram in the elevation plane is given for a frequency of \SI{5.385}{\giga\hertz}. 180° of elevation refers to the SMA-Connector side and 0° to the upper edge of the pcb antenna.}
\end{figure}

The lab setup we investigate is shown in Fig.~\ref{fig:WholeSetup} and Fig.~\ref{fig:SplitViewA}. The RIS is mounted vertically on a flat office desk in a typical office space of \SI{18}{\square\meter}.

In a distance of \SI{100}{\centi\meter} to the surface, two vertically polarized planar elliptical dipole ultra wideband antennas, specifically Broadspec\textsuperscript{TM} antennas of manufacturer Time Domain, now Humatics, \cite{TimeDomain} are placed. The two dipole antennas are used as the transmitting and receiving antenna interchangeably. They connect to the VNA or SDR with \SI{1}{\meter} RG-142 cables. The normalized antenna characteristic of the antennas at the carrier frequency of \SI{5.385}{\giga\hertz} used during the experiments is shown in Fig.~\ref{fig:SplitViewB}. As can be seen from the diagram at an angle of 180°, the antenna' pattern is attenuated by approx.~\SI{22}{\decibel}. Thus, it is a good choice to place both antennas facing outwards, as shown in Fig.~\ref{fig:SplitViewA}. The transmit and receive antenna therefore exploit the antenna pattern to yield good initial analog isolation, while keeping the antennas SMA-connectors at a distance of $\SI{25}{\milli\meter}$, which is approximately $\lambda/2$. This is a realistic spacing for mounting the antennas on a regular UE.

Both the VNA or SDR and the RIS connect to a host computer running LabView ensuring proper synchronization in between switching of the RIS and measurement of the channel.

\section{Experiment results}\label{sec:ExpResults}
In this section, we present the experimental results of the conducted case study.

\subsection{Antenna initial isolation}
\begin{figure}
	\centering
	\includegraphics[width=\linewidth]{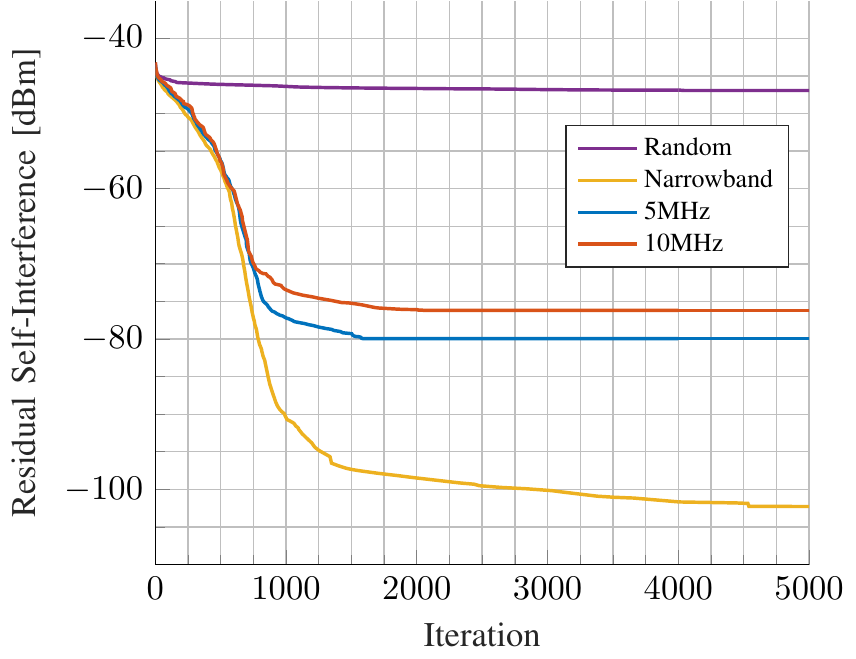}
	\caption{Convergence plot of the greedy algorithm on the lab setup. The shown numbers are averaged from 100 runs of the greedy algorithm for each case. The noise floor of the VNA is \SI{-124}{\dBm/\hertz} in the evaluated frequency band.}
	\label{fig:Convergence}
\end{figure}

\begin{figure}
	\centering
	\includegraphics[width=\linewidth]{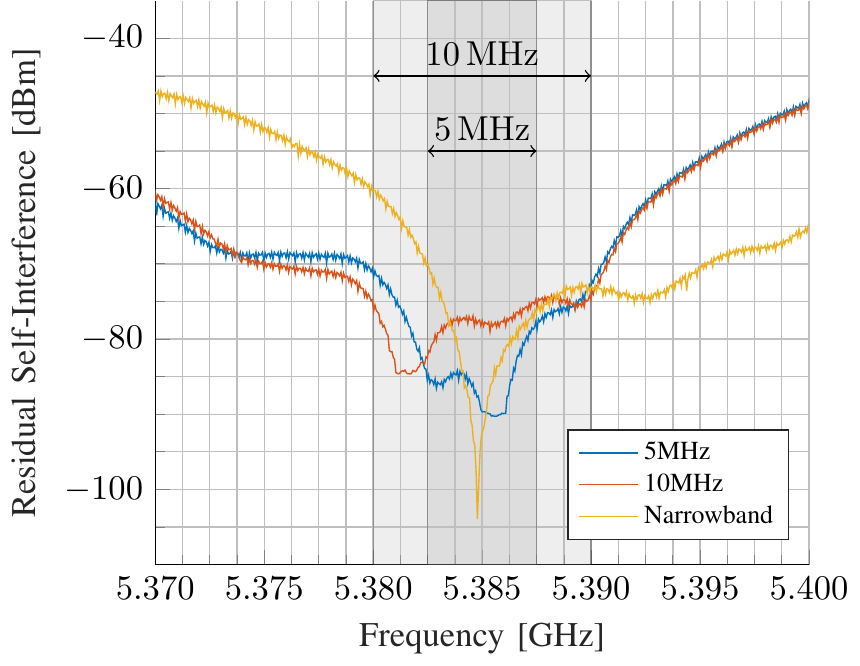}
	\caption{Measured transfer function at the antenna from transmitter to the receiver for the best RIS configuration found with the greedy algorithm.}
	\label{fig:Bandwidth}
\end{figure}

To determine the baseline performance of the antenna assembly in realistic indoor scenarios with multi-path reflections, we conduct a brief field trial. The antenna assembly is hooked up to the VNA and is moved around in the office space on a lab cart. 
 While the transmission parameter $s_{21}$ between the antennas is measured 2 times per second, the cart is pushed around on a randomized trajectory.
 In total 1200 samples are collected within \SI{10}{\minute}. We found the average leakage from the transmit to the receive antenna to be around \SI{-44}{\decibel}, with a maximum reached at \SI{-32}{\decibel} for rare cases. This gives us a baseline performance for the analog isolation of approximately \SI{44}{\decibel} provided by the antenna setup alone, which we aim to further reduce by utilizing the RIS.

\subsection{RIS assisted SI cancellation}
With the RIS separated by \SI{1}{\meter} from the antenna assembly, as shown in Fig.~\ref{fig:WholeSetup}, the residual SI is measured by evaluating the transmission parameter $s_{21}$.
First, the RIS is configured with plain random configurations, while the best SI magnitude value is kept in the buffer. The cumulative minimum of this approach, indicating the highest SI cancellation, is shown in the convergence plot in Fig.~\ref{fig:Convergence} as the graph 'Random'. It is visible that by guessing random configurations of the RIS no significant improvement is achieved. The minimal SI magnitude reached with this approach is \SI{-46}{\decibel}.
That is why we propose to use a greedy approach, as described in section~\ref{sec:Methodology}. The setup is kept exactly the same, but this time our greedy algorithm is used to optimize the RIS setting. The results are shown in Fig.~\ref{fig:Convergence} 'Narrowband'. The cumulative minimum of our proposed algorithm converges within approximately 1500~iterations to a very low residual SI magnitude of \SI{-95}{\dBm}, which further decreases with more iterations to \SI{-103}{\dBm}.

Since our algorithm has a stochastic component, the numbers shown in Fig.~\ref{fig:Convergence} are averages of a total of 100~runs of the greedy algorithm to obtain meaningful data. Each run of the algorithm is run until no improvement is found for $t_e$~iterations. Fig.~\ref{fig:Convergence} shows the averaged values for the first 5000 iterations. In our current LabView implementation with the USRPs, we reach a loop rate of \SI{25}{\hertz}, which is mainly limited by the host-driven setup and interface latency. Converging to a good SI cancellation therefore is achieved under a minute with potential for improvement. The utilizable bandwidth for the best setting found by our algorithm is given in Fig.~\ref{fig:Bandwidth} 'Narrowband'. The algorithm optimizes the residual SI at the center frequency of \SI{5.385}{\giga\hertz}. It is visible for the narrowband case that the best cancellation is achieved at a very small bandwidth, which is expected since our algorithm optimizes to a single frequency.

To broaden the usable bandwidth for, e.g., an OFDM signal we modify our algorithm to use a set of equidistant frequency points within a desired bandwidth. From all considered frequency points in every iteration the highest SI magnitude is used for the next greedy steps.
This ensures, that the residual SI magnitude is at least the converged value over the desired bandwidth.
We evaluate this approach for \SI{5}{\mega\hertz} and \SI{10}{\mega\hertz} of desired bandwidth. In Fig.~\ref{fig:Convergence}, we can see that for the \SI{5}{\mega\hertz} and \SI{10}{\mega\hertz} case, the algorithm is still able to add a significant amount of SI cancellation over the purely random case. However, the lowest achievable SI is \SI{-76}{\dBm} for a \SI{10}{\mega\hertz} band and \SI{-80}{\dBm} for a \SI{5}{\mega\hertz} band. Fig.~\ref{fig:Bandwidth} shows the corresponding transfer function, where it is visible, that the bandwidth constraint is met. The \SI{5}{\mega\hertz} band and \SI{10}{\mega\hertz} band are layed out symmetrical around the center frequency of \SI{5.385}{\giga\hertz}. Within the band the SI magnitude is at least the convergence value given in Fig.~\ref{fig:Convergence}, but often is lower in the considered interval. We can conclude that we can broaden the bandwidth in exchange for higher SI magnitude. However, the cancellation effect of the RIS is still high with an additional \SI{30}{\decibel} cancellation over the case without the RIS for \SI{10}{\mega\hertz} of bandwidth.

All described measurements were conducted two times, first with the VNA and later on with the X310 to speed up the implementation moving towards a real-time application of the concept.
We were able to reproduce the measurements with the USRPs at \SI{5,385}{\giga\hertz}, while holding the gain setting of them fixed. For brevity and accuracy, we only show the results of the VNA.

\section*{Conclusion}

In this paper, we have shown that SI cancellation for FD communication systems can be significantly enhanced by utilizing an RIS. We conducted a case study on a prototype system, which showed very promising performance in a lab scenario. Furthermore, we have proposed a greedy algorithm to generate an optimized RIS configuration in reasonable time. In addition, we modified the proposed algorithm to allow higher bandwidth and have shown that in exchange for some SI cancellation we can broaden the utilizable bandwidth. The presented approach seems to be very promising with the potential to save costs in FD transceivers and reduce the complexity of analog transceiver hardware, possibly even adopt commodity hardware in the future.

%---------------------------------------------------------------------------------------
%	REFERENCES
%----------------------------------------------------------------------------------------

\bibliographystyle{IEEEtran}
\bibliography{./bib} 

\begin{thebibliography}{10}
\providecommand{\url}[1]{#1}
\csname url@samestyle\endcsname
\providecommand{\newblock}{\relax}
\providecommand{\bibinfo}[2]{#2}
\providecommand{\BIBentrySTDinterwordspacing}{\spaceskip=0pt\relax}
\providecommand{\BIBentryALTinterwordstretchfactor}{4}
\providecommand{\BIBentryALTinterwordspacing}{\spaceskip=\fontdimen2\font plus
\BIBentryALTinterwordstretchfactor\fontdimen3\font minus
  \fontdimen4\font\relax}
\providecommand{\BIBforeignlanguage}[2]{{%
\expandafter\ifx\csname l@#1\endcsname\relax
\typeout{** WARNING: IEEEtran.bst: No hyphenation pattern has been}%
\typeout{** loaded for the language `#1'. Using the pattern for}%
\typeout{** the default language instead.}%
\else
\language=\csname l@#1\endcsname
\fi
#2}}
\providecommand{\BIBdecl}{\relax}
\BIBdecl

\bibitem{cisco}
\BIBentryALTinterwordspacing
{Cisco Systems, Inc.} Cisco annual internet report (2018-2023). [Online].
  Available:
  \url{https://www.cisco.com/c/en/us/solutions/collateral/executive-perspectives/annual-internet-report/white-paper-c11-741490.pdf}
\BIBentrySTDinterwordspacing

\bibitem{zhao2019survey}
J.~{Zhao}, ``{A Survey of Intelligent Reflecting Surfaces (IRSs): Towards 6G
  Wireless Communication Networks},'' \emph{arXiv e-prints}, p.
  arXiv:1907.04789, Jul. 2019.

\bibitem{PStaat}
P.~Staat, H.~Elders-Boll, M.~Heinrichs, R.~Kronberger, C.~Zenger, and C.~Paar,
  ``Intelligent reflecting surface-assisted wireless key generation for
  low-entropy environments,'' 2021, pre-print; accepted in IEEE 32nd Annual
  International Symposium on Personal, Indoor and Mobile Radio Communications
  (PIMRC).

\bibitem{SecureMIMO}
L.~Dong and H.-M. Wang, ``Enhancing secure mimo transmission via intelligent
  reflecting surface,'' \emph{IEEE Transactions on Wireless Communications},
  vol.~19, no.~11, pp. 7543--7556, 2020.

\bibitem{AntiJamming}
H.~Yang, Z.~Xiong, J.~Zhao, D.~Niyato, Q.~Wu, H.~V. Poor, and M.~Tornatore,
  ``Intelligent reflecting surface assisted anti-jamming communications: A fast
  reinforcement learning approach,'' \emph{IEEE Transactions on Wireless
  Communications}, vol.~20, no.~3, pp. 1963--1974, 2021.

\bibitem{Katti1}
\BIBentryALTinterwordspacing
M.~Jain, J.~I. Choi, T.~Kim, D.~Bharadia, S.~Seth, K.~Srinivasan, P.~Levis,
  S.~Katti, and P.~Sinha, ``Practical, real-time, full duplex wireless,'' in
  \emph{Proceedings of the 17th Annual International Conference on Mobile
  Computing and Networking}, ser. MobiCom '11.\hskip 1em plus 0.5em minus
  0.4em\relax New York, NY, USA: Association for Computing Machinery, 2011, p.
  301–312. [Online]. Available: \url{https://doi.org/10.1145/2030613.2030647}
\BIBentrySTDinterwordspacing

\bibitem{Katti2}
\BIBentryALTinterwordspacing
J.~I. Choi, M.~Jain, K.~Srinivasan, P.~Levis, and S.~Katti, ``Achieving single
  channel, full duplex wireless communication,'' in \emph{Proceedings of the
  Sixteenth Annual International Conference on Mobile Computing and
  Networking}, ser. MobiCom '10.\hskip 1em plus 0.5em minus 0.4em\relax New
  York, NY, USA: Association for Computing Machinery, 2010, p. 1–12.
  [Online]. Available: \url{https://doi.org/10.1145/1859995.1859997}
\BIBentrySTDinterwordspacing

\bibitem{Duarte1}
M.~Duarte, C.~Dick, and A.~Sabharwal, ``Experiment-driven characterization of
  full-duplex wireless systems,'' \emph{IEEE Transactions on Wireless
  Communications}, vol.~11, no.~12, pp. 4296--4307, 2012.

\bibitem{Duarte2}
M.~Duarte and A.~Sabharwal, ``Full-duplex wireless communications using
  off-the-shelf radios: Feasibility and first results,'' in \emph{2010
  Conference Record of the Forty Fourth Asilomar Conference on Signals, Systems
  and Computers}, 2010, pp. 1558--1562.

\bibitem{FDradios}
\BIBentryALTinterwordspacing
D.~Bharadia, E.~McMilin, and S.~Katti, ``Full duplex radios,'' \emph{SIGCOMM
  Comput. Commun. Rev.}, vol.~43, no.~4, p. 375–386, Aug. 2013. [Online].
  Available: \url{https://doi.org/10.1145/2534169.2486033}
\BIBentrySTDinterwordspacing

\bibitem{Vogt}
H.~Vogt, G.~Enzner, and A.~Sezgin, ``State-space adaptive nonlinear
  self-interference cancellation for full-duplex communication,'' \emph{IEEE
  Transactions on Signal Processing}, vol.~67, no.~11, pp. 2810--2825, 2019.

\bibitem{Everett}
E.~Everett, C.~Shepard, L.~Zhong, and A.~Sabharwal, ``Softnull: Many-antenna
  full-duplex wireless via digital beamforming,'' \emph{IEEE Transactions on
  Wireless Communications}, vol.~15, no.~12, pp. 8077--8092, 2016.

\bibitem{WCTRIS}
E.~Basar, M.~Di~Renzo, J.~De~Rosny, M.~Debbah, M.-S. Alouini, and R.~Zhang,
  ``Wireless communications through reconfigurable intelligent surfaces,''
  \emph{IEEE Access}, vol.~7, pp. 116\,753--116\,773, 2019.

\bibitem{Heinrichs}
M.~Heinrichs and R.~Kronberger, ``Digitally tunable frequency selective surface
  for a physical layer security system in the 5 ghz wi-fi band,'' \emph{2020
  International Symposium on Antennas and Propagation (ISAP)}, pp. 267--268,
  2021.

\bibitem{agilent}
\BIBentryALTinterwordspacing
{Keysight Technologies Inc. formerly Agilent Technologies}. Datasheet e5071c
  ena vector network analyzer; 5989-5479. [Online]. Available:
  \url{https://www.keysight.com/de/de/assets/7018-01424/data-sheets/5989-5479.pdf}
\BIBentrySTDinterwordspacing

\bibitem{X310}
\BIBentryALTinterwordspacing
{Ettus Research}. Specifications x310 software defined radios. [Online].
  Available: \url{https://kb.ettus.com/X300/X310}
\BIBentrySTDinterwordspacing

\bibitem{CBX}
\BIBentryALTinterwordspacing
------. Specifications cbx daughterboard. [Online]. Available:
  \url{https://kb.ettus.com/CBX}
\BIBentrySTDinterwordspacing

\bibitem{TimeDomain}
\BIBentryALTinterwordspacing
{Time Domain Inc.} (2017) Broadspec tm uwb antenna 320-0385c datasheet.
  [Online]. Available: \url{https://www.humatics.com}
\BIBentrySTDinterwordspacing

\end{thebibliography}

\end{document}